\documentclass{iau} 
\usepackage{graphicx}
\usepackage{amsmath}
\def\kms{\,{\rm km\,s^{-1}}}

\title[RAVE-Gaia] 
{RAVE-Gaia and the impact on Galactic archeology}

\author[Andrea Kunder \& RAVE Collaboration]   
{Andrea Kunder$^1$,$^2$ + RAVE Collaboration}

\affiliation{$^1$Leibniz-Institut f\"{u}r Astrophysik Potsdam (AIP), An der Sternwarte 16, D-14482 Potsdam, Germany \\ email: {\tt amkunder@gmail.com} \\[\affilskip]
$^2$Saint Martin's University, 5000 Abbey Way SE, Lacey, WA 98503, USA }

\pubyear{2017}
\volume{330}  
\jname{Astrometry and Astrophysics in the Gaia Sky}
\editors{A. Recio-Blanco \& P. de Laverny, eds.}
\begin{document}

\maketitle

\begin{abstract}
The new data release (DR5) of the RAdial Velocity Experiment (RAVE) includes radial velocities of 
520,781 spectra of 457,588 individual stars, of which 215,590 individual stars are released in the 
Tycho-{\it Gaia} astrometric solution (TGAS) in {\it Gaia} DR1. Therefore, RAVE contains the largest TGAS 
overlap of the recent and ongoing Milky Way spectroscopic surveys. Most of the RAVE stars also 
contain stellar parameters (effective temperature, surface gravity, overall metallicity), as well as 
individual abundances for Mg, Al, Si, Ca, Ti, Fe, and Ni. Combining RAVE with TGAS brings the 
uncertainties in space velocities down by a factor of 2 for stars in the RAVE volume -- 10 km~s$^{-1}$ 
uncertainties in space velocities are now able to be derived for the majority (70\%) of the 
RAVE-TGAS sample, providing a powerful platform for chemo-dynamic analyses of the Milky Way. 
Here we discuss the RAVE-TGAS impact on Galactic archaeology as well as how the {\it Gaia} 
parallaxes can be used to break degeneracies within the RAVE spectral regime for an even better 
return in the derivation of stellar parameters and abundances.
\keywords{astronomical data bases: RAVE, surveys, stars: kinematics, stars: abundances, 
stars: Hertzsprung-Russell diagram, Galaxy: kinematics and dynamics, 
Galaxy: stellar content, Galaxy: structure   
}
\end{abstract}

\firstsection 
\section{Introduction}
Our Milky Way galaxy contains stars that are distinctly closer and brighter to us than stars in
neighbouring galaxies, so the level of detail with which the stellar populations in our Galaxy 
can be seen provide important information regarding the formation and evolution 
of large spiral galaxies.  The motions of stars combined with their chemical abundances 
in particular place powerful constraints on the formation of spiral galaxies such as the 
Milky Way (e.g.,  \cite[Minchev, Chiappini \& Martig 2013]{minchev13}).
Today, the astrometric satellite {\it Gaia} is providing its first measurements (Data 
Release 1, {Gaia} Collaboration et~al.\ 2016), and the Tycho-{\it Gaia} Astrometric 
Solution (TGAS, \cite[Lindegren et~al.\ 2016]{lindegren16}) contains 
positions, parallaxes, and proper motions for about 2 million of the brightest stars 
in common with the Hipparcos and Tycho-2 catalogues.  With typical accuracies of 
$\sim$1 mas~yr$^{-1}$ and 0.3 mas in proper motion and parallax, respectively, this is 
comparable to the precision of Hipparcos, but on a sample that is more than an order 
of magnitude larger.  

In TGAS, exquisite astrometry is given in the positions and proper 
motions of stars.  Combined with external spectroscopy, the measure of stellar atmospheric 
parameters, individual chemical abundances and radial velocities allow a full definition of  
the motion of stars in the Galaxy.  Among existing spectroscopic surveys, the Radial Velocity 
Experiment (\cite[RAVE, Steinmetz et~al. 2006, Zwitter et al. 2008, Siebert et al. 2011,
Kordopatis et~al. 2013, Kunder et~al. 2017]{steinmetz06, zwitter08, siebert11, kordopatis13, kunder17}) 
has the largest overlap with TGAS ($>$200,000) 
so is a particularly attractive database for astronomers seeking to simultaneously use chemical 
and dynamical information to complement the available {\it Gaia} astrometry. 

\section{RAVE Overview}
RAVE is a magnitude-limited survey of stars randomly selected in the 9 $< I <$ 12 magnitude range, 
obtained from spectra with a resolution of R$\sim$7\,500 covering the CaT regime.  It currently containts 
the largest spectroscopic sample of stars in the Milky Way which overlaps with 
the {\it Gaia}-TGAS proper motions and parallaxes (Table~\ref{tab1}).
 
\begin{table}
  \begin{center}
  \caption{Overlap of large spectroscopic surveys with {\it Gaia}-TGAS.}
  \label{tab1}
 {\scriptsize
  \begin{tabular}{l|cl}\hline 
{\bf Survey} & {\bf Number TGAS stars } \\ \hline
RAVE DR5 & 215,600 \\
LAMOST DR2 & 124,300 \\
GALAH DR1 & 8,500 \\
APOGEE DR13 & 21,700 \\ \hline
  \end{tabular}
  }
 \end{center}
\vspace{1mm}
\end{table}

Radial velocities are available for all RAVE stars, where
the typical signal-to-noise (SNR) ratio of a RAVE
star is 40 and the typical uncertainty in radial velocity is $<2\kms$.
For a subsample of RAVE stars, stellar parameters are also provided.
These temperatures, $T_{\rm eff}$, gravities, $\log g$, and metallicities, $\rm [M/H]$, 
are obtained using the DR4 stellar parameter pipeline, which is built
on the algorithms of MATISSE and DEGAS, with an updated calibration that improves
the accuracy of especially the $\log g$ values of stars. The uncertainties
vary with stellar population and SNR, but for the most reliable stellar parameters,
the uncertainties in $T_{\rm eff}$,  $\log g$, and $\rm [M/H]$ are approximately 250\,K, 0.4\,dex and
0.2\,dex, respectively.  RAVE stars with the most reliable stellar parameters 
are those which have {\tt Algo\_Conv}=0 (meaning the stellar parameter algorithm converged), 
SNR $>$ 40, and {\tt c1}=n, {\tt c2}=n and {\tt c3}=n. (which means the star has
a spectrum that is ``normal").  Error spectra computed for each 
observed spectrum is used to assess the uncertainties in the radial velocities 
and stellar parameters.

The elemental abundances of Al, Si, Ti, Fe, Mg and Ni are derived for
$\sim$2/3 of the RAVE stars, which have uncertainties of $\sim0.2\,$dex, although their
accuracy varies with SNR and, for some elements, also of the stellar
population.  Distances, ages, masses and the interstellar extinctions are
computed using an upgraded method of what is presented in \cite[Binney et~al.(2014)]{binney14}.

RAVE DR5 further provides temperatures from the Infrared Flux Method,
which are available for $>95$\% of all
RAVE stars.  For a sub-sample of stars that can be calibrated
asteroseismically ($\sim45$\% of the RAVE sample), an asteroseismically
calibrated $\log g$, as detailed in \cite[Valentini et~al.(2017)]{valentini17} is provided.  
Stellar parameters of the RAVE stars are also found using the data-driven 
approach of {\it The Cannon} (\cite[Casey et~al. 2017]{casey17}), for which
$T_{\rm eff}$, surface gravity $\log{g}$ and $\rm [Fe/H]$, as well as
chemical abundances of giants of up to seven elements (O, Mg, Al, Si, Ca, Fe, Ni)
is presented.  

All of the above described information is publicly available, and can be downloaded
via the RAVE Web site {\tt http://www.rave-survey.org} or the Vizier database.

\begin{figure}[b]
\begin{center}
 \includegraphics[width=3.4in]{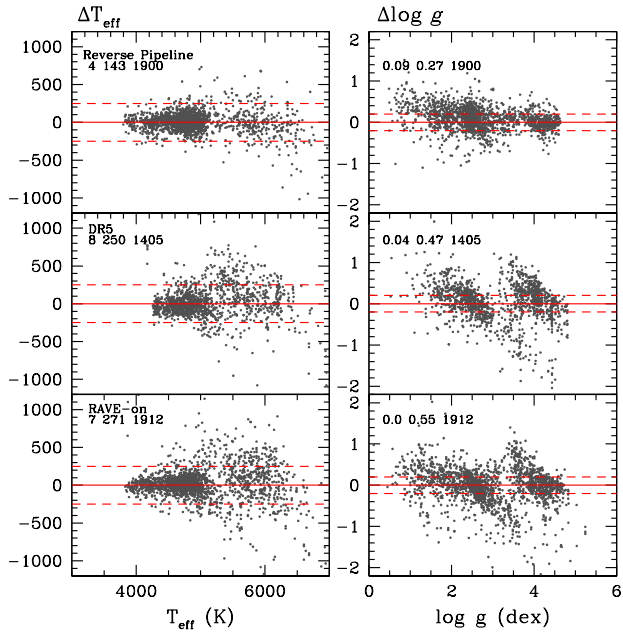} 
 \caption{The reverse distance pipeline (top), DR5 main (middle), RAVE-on (bottom) temperatures 
 and gravities compared to RAVE stars that overlap with high-resolution studies.  
 The top left corner indicates the bias, dispersion and number of stars for each comparison.
For the DR5 comparison, only stars with AlgoConv = 0 are shown. 
}
\label{fig1}
\end{center}
\end{figure}

\section{Reverse Pipeline}
It is well-known that stellar spectra with a resolution $R <$10~000 suffer from spectral
degeneracies at the Calcium triplet wavelength range.  Specifically, at the RAVE wavelength and resolution, 
hot dwarf and cool giant stars share the same spectral signatures
-- see for example Figure~1 in \cite[Matijevi\^{c} et~al.~(2017)]{matijevic17}.  
Parameter degeneracy is usually less severe when the available information about the 
parameters increases: e.g., with a wider spectral range, higher spectral resolution, etc.
The TGAS parallaxes can provide powerful extra information to break 
degeneracies, thereby constraining stellar parameters.  

The RAVE distance pipeline (as described in Binney et~al. 2014 and Kunder et~al. 2017) 
takes as its input $\rm T_{eff}$, $\log g$, $\rm [M/H]$, $J$, $H$ and $K$ magnitudes
and with this information combined with stellar isochrones, descriptions of the posterior 
probabilities of different properties of the stars (e.g., mass, age, line-of-sight extinction, 
distance) are generated.  It has been modified to now also take 
the TGAS parallaxes, as well as AllWISE $W1$ and $W2$ magnitudes as an input 
(McMillan et~al. 2017, in prep).  Using the same prior as in Binney et~al. (2014), a new $\log g$, 
$\rm T_{eff}$ and $\rm [M/H]$ is found, and for the first time, descriptions of the posterior 
probabilities for $\rm T_{eff}$, $\log g$ and $\rm [M/H]$ are obtained.  We therefore 
refer to this as
the `reverse pipeline', because rather than just taking the stellar parameters as input,
they are an end product.  In fact, these are an inevitable byproducts of the distance pipeline, 
produced because each "model star" is compared to the data which has an 
associated $\rm T_{eff}$, $\log g$ and $\rm [M/H]$ as the likelihood is calculated.

Figure~\ref{fig1} shows how the reverse distance pipeline (top), DR5 main (middle), 
RAVE-on (Casey et al. 2017) (bottom) temperatures and gravities compare to RAVE stars that fortuitously 
overlap with high-resolution studies (e.g., Gaia-ESO, globular and open clusters, 
GALAH and field star surveys -- see DR5 paper for details). The reverse pipeline yields temperatures and gravities 
that agree better to external, high-resolution studies of RAVE stars than both DR5 and RAVE-on. 
Note that the reverse pipeline temperatures and gravities are only available for TGAS stars.

The largest discrepancies between the DR5 and reverse pipeline main temperatures and 
gravities occur at the giant/dwarf interface.  This is expected, as this is where the degeneracies 
mentioned above are the most severe.  Our preliminary tests show no signs of bias in the reverse 
pipeline stellar parameters as a function of parallax or TGAS parallax uncertainty.  We are 
carrying out extensive tests to check if any and what kinds of subtle biases may exist when 
applying the reverse pipeline.

\begin{figure}
\includegraphics[width=2.8in]{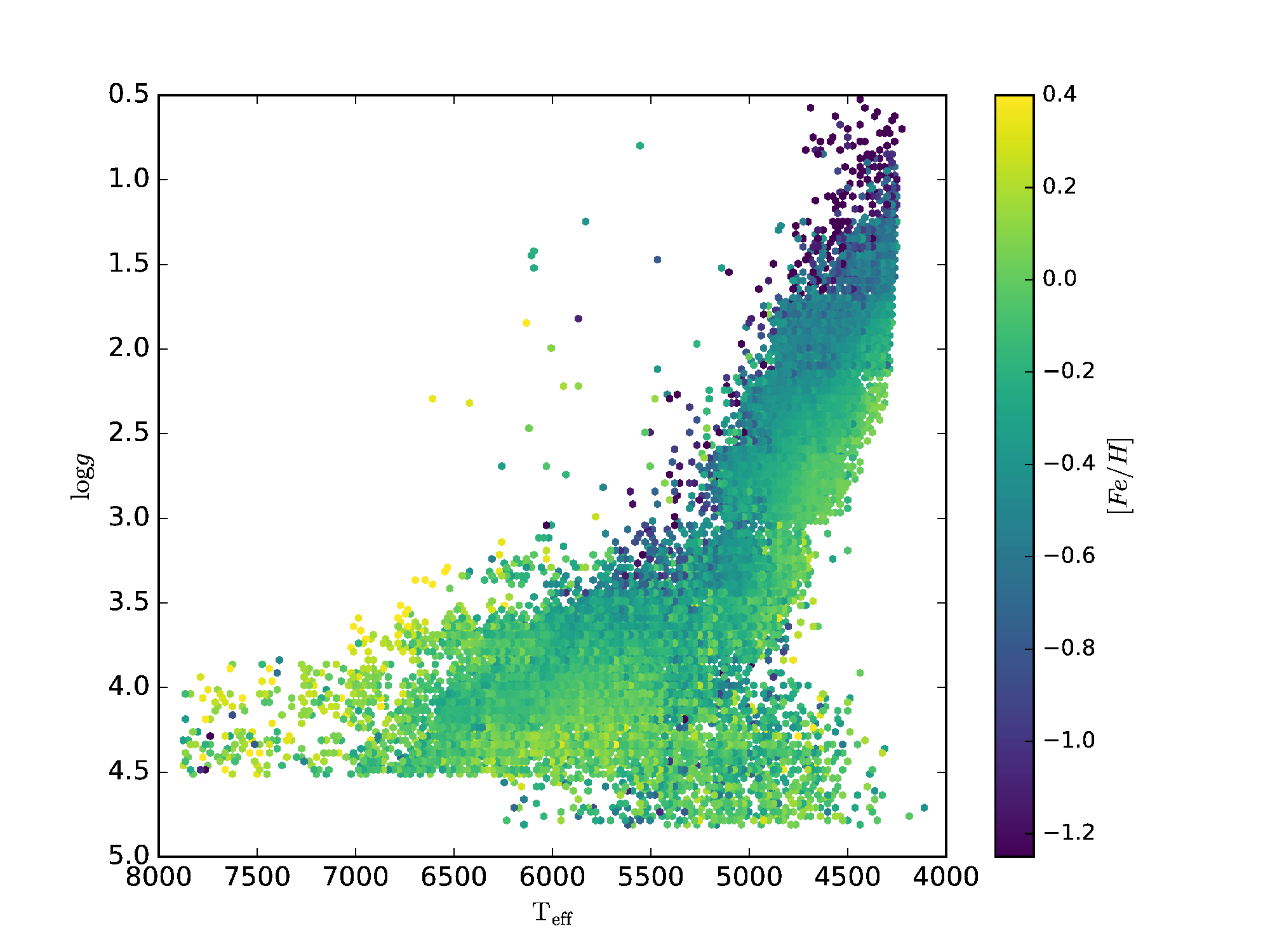}
\includegraphics[width=2.8in]{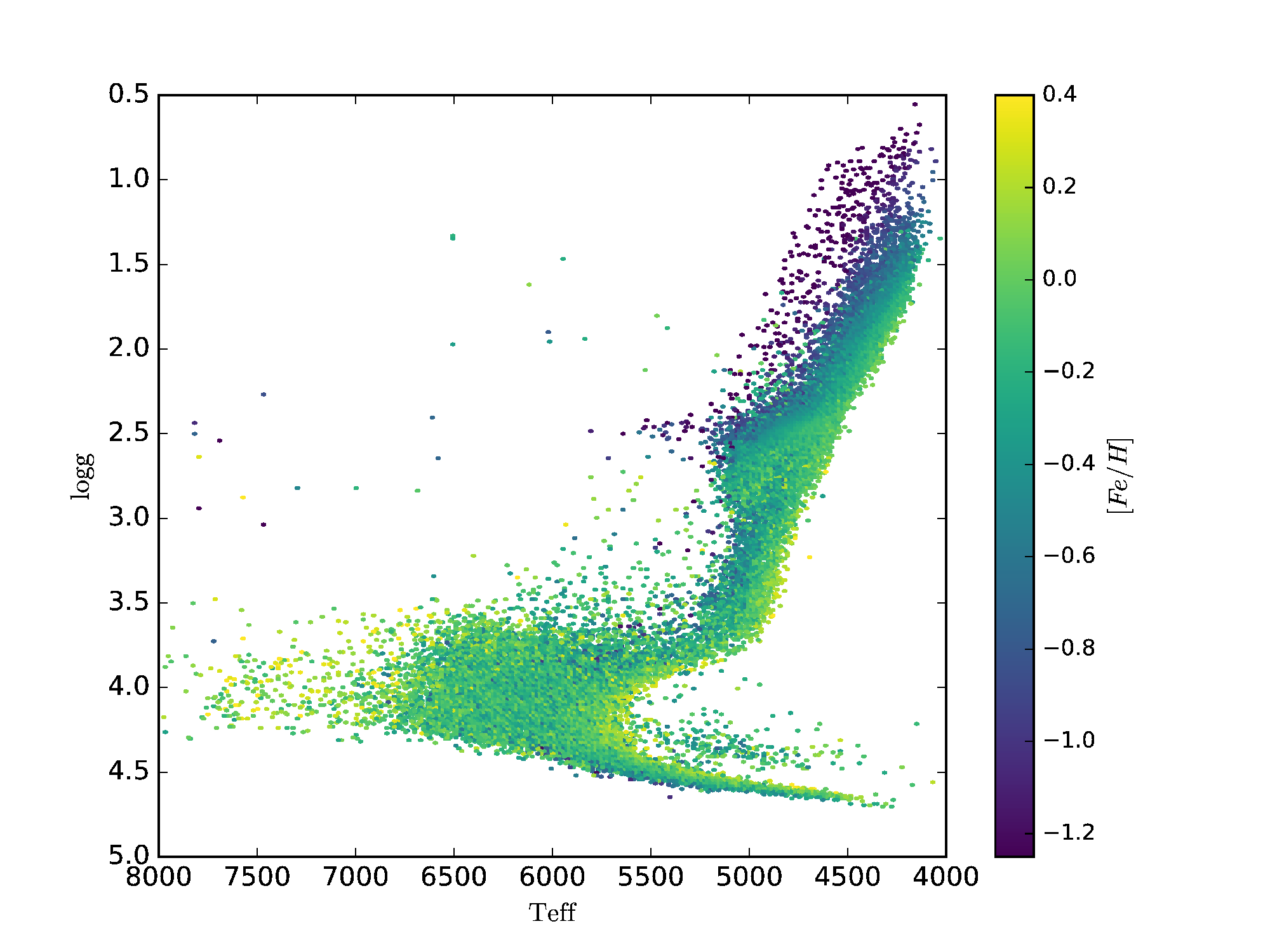}
\caption{$T_{\rm eff}$-$\log g$ diagram for calibrated DR5 parameters (left)
and reverse pipeline parameters (right). 
\label{figHRD}}
\end{figure}

Figure~\ref{figHRD} (right) shows the on the Hertzsprung Russel diagram using the DR5 
and reverse distance pipeline temperatures and gravities. Of particular interest is the
narrow main sequence, and a sequence of stars above the main-sequence
separated by a clear gap.  The majority of these stars are double-lined spectroscopic binaries (SB2s) 
which (in the absence of eclipses) are not variable 
and where the orbital period is short enough to permit any astrometric signature.  
Hence, they do not fall into photometrically or astrometrically peculiar classes, and are
included in {\it Gaia} DR1. 
This binary main-sequence is a consequence of the fact that nature perfers to make binaries with 
a mass ratio close to 1 -- which have photometry and astrometry just like single stars, 
but are 2-times ($\sim$0.75 mag) brighter. 


\section{Conclusions}
RAVE is continuing to yield exciting results using the data products {\it Gaia} DR1.  The
reverse pipeline, described above, is allowing more accurate stellar parameters to be
obtained, which can
then be fed into a new elemental abundance pipeline designed specifically for the RAVE
spectra (Guiglion et~al. 2016, Guiglion et~al. 2017, in prep).  Jofre et~al.(2017, submitted) has expanded
the number of RAVE stars with TGAS parallax uncertainties less than 20\% by applying 
the twin method to RAVE.  McMillan et~al.(2017, in prep)
is using the TGAS parallaxes to find more precise distance estimates for all the RAVE stars,
which also has the effect of an improvement in age uncertainties.  Therefore, an exploration of
the correlation between ages, metallicities, and velocities of stars in the solar neighborhood 
can be carried out (Wojno et~al. 2017, in prep).  Last but not least, 200 light curves of RAVE stars
in the K2-Campaign 6 have been analysed, which will be used
as calibration data for log~$g$ (Valentini et~al. 2017, in prep).

\section{Acknowlegements}
This work has made use of data from the European Space Agency (ESA)
mission {\it Gaia} (\url{https://www.cosmos.esa.int/gaia}), processed by
the {\it Gaia} Data Processing and Analysis Consortium (DPAC,
\url{https://www.cosmos.esa.int/web/gaia/dpac/consortium}). Funding
for the DPAC has been provided by national institutions, in particular
the institutions participating in the {\it Gaia} Multilateral Agreement.
Funding for RAVE has been provided by: the Australian Astronomical Observatory; the 
Leibniz-Institut fuer Astrophysik Potsdam (AIP); the Australian National University; the 
Australian Research Council; the French National Research Agency; the German Research 
Foundation (SPP 1177 and SFB 881); the European Research Council (ERC-StG 240271 Galactica); 
the Istituto Nazionale di Astrofisica at Padova; The Johns Hopkins University; the National Science 
Foundation of the USA (AST-0908326); the W. M. Keck foundation; the Macquarie University; the 
Netherlands Research School for Astronomy; the Natural Sciences and Engineering Research 
Council of Canada; the Slovenian Research Agency; the Swiss National Science Foundation; the 
Science \& Technology Facilities Council of the UK; Opticon; Strasbourg Observatory; and the 
Universities of Groningen, Heidelberg and Sydney.

\end{document}